\documentclass{acmart}

\usepackage{lscape}
\usepackage{longtable}

\setcopyright{none}
\begin{document}

%%
%% The "title" command has an optional parameter,
%% allowing the author to define a "short title" to be used in page headers.
\title[GVP]{Gradual Voluntary Participation: A Framework for Participatory AI Governance in Journalism}

\author{Stefano Sorrentino}
\authornote{Both authors contributed equally to the paper}
\email{stefano.sorrentino@epfl.ch}
\orcid{0009-0001-9166-9396}
\affiliation{
  \institution{École Polytechnique Fédérale de Lausanne (EPFL)}
  \city{Lausanne}
  \state{}
  \country{Switzerland}
}

\author{Matilde Barbini}
\authornotemark[1]
\email{matilde.barbini@epfl.ch}
\orcid{0009-0007-7986-2365}
\affiliation{
  \institution{École Polytechnique Fédérale de Lausanne (EPFL)}
  \city{Lausanne}
  \country{Switzerland}
}

\author{Daniel Gatica-Perez}
\email{gatica@idiap.ch}
\orcid{0000-0001-5488-2182}
\affiliation{
  \institution{Idiap Research Institute}
  \city{Martigny}
  \country{Switzerland}
}
\affiliation{
  \institution{École Polytechnique Fédérale de Lausanne (EPFL)}
  \city{Lausanne}
  \country{Switzerland}
}

%%
%% The "author" command and its associated commands are used to define
%% the authors and their affiliations.
%% Of note is the shared affiliation of the first two authors, and the
%% "authornote" and "authornotemark" commands
%% used to denote shared contribution to the research.

%% TODO: Replace with your actual author information
%\author{[Your Name]}
%\email{[your.email@institution.edu]}
%\orcid{[your-orcid-id]}
%\affiliation{%
 % \institution{[Your Institution]}
 % \city{[City]}
 % \state{[State/Province]}
 % \country{[Country]}
%}

%% Add additional authors as needed
%\author{[Co-author Name]}
%\email{[coauthor.email@institution.edu]}
%\affiliation{%
%  \institution{[Co-author Institution]}
%  \city{[City]}
%  \country{[Country]}
%}

%%
%% By default, the full list of authors will be used in the page
%% headers. Often, this list is too long, and will overlap
%% other information printed in the page headers. This command allows
%% the author to define a more concise list
%% of authors' names for this purpose.
%\renewcommand{\shortauthors}{[Author et al.]}

%%
%% The abstract is a short summary of the work to be presented in the
%% article.

\begin{abstract}

The integration of AI into journalism challenges participatory design (PD), particularly with respect to stakeholder influence, workplace perceptions, and organizational dynamics. Traditional PD assumes that users can shape technologies, yet AI systems resist influence due to opaque data, fixed architectures, and inaccessible objectives. Through interviews with 10 journalists, we identify the \textit{perception gap}, showing that trust in AI depends on perceived agency within workplace participatory workflows.
Informed by these findings, we introduce the Gradual Voluntary Participation (GVP) framework in journalism and its five core principles, reconceptualizing participation as a gradual and voluntary process that can be operationalized at the newsroom level, beyond fixed workshops or one-time preference-elicitation campaigns. Addressing epistemic burdens, participatory ceilings, and performative consultations, GVP treats gradualism and voluntariness as design dimensions that shape perception, legitimacy, and ownership. Moving beyond unidimensional ladder metaphors and adopting a bidimensional matrix structure, the framework maps stakeholders across depth and scope, offering a new model for local participatory AI governance that balances technological transformation with stakeholder empowerment in rapidly evolving hybrid workplaces.
\end{abstract}




\maketitle

%% INTRODUCTION
\section{Introduction}

Newsrooms have long been early adopters of technological change, and the integration of large language models (LLMs) into journalistic practice has opened new possibilities for experimentation \cite{sonni_digital_2024}. Yet, AI adoption in journalism%is not only a matter of introducing new tools. It also 
raises organizational questions about who can shape these systems, under what conditions, and with what consequences for editorial authority, professional accountability, and newsroom coordination \cite{diakopoulos2017algorithmic}. As AI enters high-stakes professional settings, participation becomes not only a matter of consultation or usability, but of how groups negotiate influence over technologies that affect their work while remaining only partially open to local intervention. In journalism, however, participation is related to editorial choices, professional accountability, and public-facing responsibility. Unlike other professional domains, newsroom actors operate under norms of verification, attribution, and audience trust, which make the question of who can influence AI systems both epistemic and civic, beyond purely organizational.

This creates a challenge for participatory design (PD), which has historically emphasized stakeholders’ ability to shape the technologies that affect them \cite{bannon_reimagining_2018, berditchevskaia_participatory_nodate, gregory_scandinavian_nodate, kraft_collective_1994, sundblad_utopia_2011}. With contemporary AI systems, many consequential decisions remain upstream, with participation unevenly distributed across implementation layers: training data, model architectures, optimization objectives, and update cycles are often inaccessible to the professionals asked to use these tools \cite{suresh_participation_2024}.

Existing work has shown that participation in AI can become tokenistic, burdensome, or disconnected from actual power \cite{sloane_participation_2022, birhane_power_2022}, with journalism research likewise documenting concerns around trust, autonomy, and ownership in AI adoption \cite{thasler2025journalistic, tseng_ownership_2025}. How newsroom actors evaluate the legitimacy of participation itself is less clear, i.e., how they distinguish between forms of involvement that support editorial agency and those that merely shift responsibility onto already constrained professionals. This question is especially pressing in resource-limited news organizations, where participation competes with deadlines, hierarchies, and dependence on external vendors \cite{breed_social_1955, simon_artificial_2024, simon_uneasy_2022}.

In this paper, we examine how journalists and editors in an independent, continental European newsroom\footnote{The specific country of the newsroom is omitted here to preserve double-blind review.} understand and evaluate participation in AI adoption. Drawing on semi-structured interviews with 10 newsroom actors, we show that participants did not primarily evaluate AI integration through performance or efficiency claims, but through the conditions under which AI would enter newsroom practice: whether editorial sovereignty was preserved, responsibilities remained accountable, systems became intelligible in use, and participation was recognized as legitimate work rather than additional burden. We describe this empirical pattern as a \emph{perception gap}: a mismatch between formal and nominal means of participation in AI adoption and the organizational conditions under which participation is experienced as meaningful.

From these findings, we develop \emph{Gradual Voluntary Participation} (GVP), a framework for participatory AI governance in journalism. GVP conceptualizes participation across two dimensions: \emph{depth}, the degree of influence actors have over AI-related decisions and arrangements, and \emph{scope}, the arenas in which participation takes place, from local tool use to broader organizational and external governance contexts. It adopts a matrix form to make participation visually concrete while framing it as conditional, non-linear, and shaped by the practical realities of newsroom work. The framework is articulated through five principles derived from our data: recognition without coercion, continuous value alignment, contextual knowledge scaffolding, sovereignty preservation, and transparent accountability.

Our contribution is threefold. First, we provide an empirical account of how newsroom actors evaluate participation in AI adoption through questions of agency, accountability, and organizational credibility rather than technical performance alone. Second, we introduce GVP as a matrix-based framework that moves beyond static, unidimensional "ladder-like" models of participation \cite{arnstein_ladder_1969} by conceptualizing participation across organizational depth and collective scope in AI-integrated newsroom settings. Third, we articulate a set of principles for supporting participatory AI governance under conditions of workload pressure, role asymmetry, and limited leverage over external AI infrastructures.

By reframing AI adoption as a problem of organizational participation and coordination, %rather than tool acceptance alone, 
this paper contributes to ongoing discussions in participatory design and human-AI interaction about how professional groups can exercise meaningful influence over opaque and externally governed systems.

\section{Related Work: Participation, Power, and Organizational Reconfiguration in AI-Mediated Journalism}
\label{perceptiongap}

Participatory design (PD) has long treated the expansion of stakeholder \emph{voice} and \emph{power} as a democratic aspiration, especially where technologies reshape work practices and decision-making relations \cite{bannon_reimagining_2018, cooper_systematic_2022, cooper_fitting_2024, delgado_participatory_2023}. Yet participatory outcomes are difficult to evaluate in heterogeneous organizational settings, particularly when what is being designed is not a bounded artifact but an evolving sociotechnical arrangement embedded in ongoing work \cite{bratteteig_what_2016, delgado_participatory_2023, suresh_participation_2024}. These difficulties intensify in AI contexts, where retraining, adaptation, and changing data inputs make participation hard to locate at any single design moment \cite{bratteteig_does_2018, bratteteig_what_2016}. Participation therefore cannot be assessed only by whether stakeholders are consulted, but also by who can influence implementation, which decisions remain contestable, and how authority is distributed over time \cite{delgado_participatory_2023, feffer_preference_2023}. Recent work on "participation in the age of foundation models" \cite{suresh_participation_2024} conceptualizes AI systems as stratified across "foundation", "subfloor", and "surface" layers, showing how participatory interventions tend to cluster at the interface level, while excluding stakeholders from upstream decisions about data, optimization, and infrastructure. In journalism, this stratification maps directly onto newsroom-tool relationships, where journalists primarily engage with interfaces while relying on external actors for the underlying system design and development, reinforcing structural limits to meaningful participation \cite{suresh_participation_2024}.

This problem is especially acute in newsrooms, framed as hierarchical organizations in which editors and publishers set policies and control resources, while reporters exercise uneven and status-dependent autonomy \cite{breed_social_1955, pavlik_impact_2000}. Editorial independence, while central to journalistic identity, is therefore conditional on funding arrangements, infrastructural access, organizational relationships, and technological dependencies \cite{drunen_safeguarding_2023, paik_journalism_2025, simon_uneasy_2022, simon_escape_2024}. Participation in journalistic AI should not be framed simply as stakeholder inclusion in design, but as negotiating influence within settings already structured by unequal decision rights, professional norms, and external constraints \cite{cooper_fitting_2024, corbett_power_2023, tseng_ownership_2025}.

\paragraph{Journalism as an organizational site of power.}
Both recent and foundational scholarship shows that in complex organizations such as newsrooms, power operates through direct control, agenda-setting, preference-shaping, and the normalization of certain assumptions and discourses over others \cite{clegg_project_2024, dahl_1957_concept, dundon2017power, jungherr_discursive_2019}. This matters for AI because design and deployment do not introduce tools into neutral settings; they can also encode, formalize, and stabilize existing institutional logics \cite{gitelman2013raw, navigli_biases_2023, neumann_position_2025}. Workflow technologies act not only as coordination mechanisms but also as organizational \emph{accounting devices} that render work visible, allocable, and subject to managerial oversight \cite{dourish2001process}. In journalism, where editorial authority is already contested across managerial, commercial, technical, and professional lines \cite{breed_social_1955, drunen_safeguarding_2023, pavlik_impact_2000}, questions of participation are inseparable from questions of who defines the goals of AI adoption, whose expertise counts as legitimate, and which values become embedded in systems and workflows \cite{feffer_preference_2023, gerdes_participatory_2022, neghawi_linking_2023, tseng_ownership_2025}.

\paragraph{AI as a reconfiguration of intra-newsroom group dynamics.}
AI adoption redistributes power and editorial authority within the newsroom: responsibilities once associated with editors and reporters may shift toward managers, technical specialists, and algorithmic systems, while journalists become more tightly coupled to digitized workflows that can narrow their discretion \cite{dodds2024impact, garcia_aviles_journalists_2004, kung2013innovation, thasler2025journalistic}. Yet the effects of algorithmic governance are not uniformly coercive. Ethnographic work on platform-based algorithmic management suggests that workers may retain significant discretion even under opaque control, depending on organizational context and the specific configuration of human-algorithmic authority \cite{kusk2022working}. At the same time, separation between editorial and engineering units fosters knowledge silos and weak cross-departmental coordination, making it harder for non-technical actors to contest system behavior or locate accountability \cite{cools_news_2024, dodds2024impact, dorr2017ethical}. Research on groupware adoption in distributed organizations similarly show how conflicting views of collaboration, incompatible standards, and uneven access to resources shape which groups can appropriate shared tools and which remain excluded \cite{mark2003shaping}. These dynamics also produce hybrid roles that may open new possibilities for collaboration while unsettling existing boundaries of expertise and control \cite{albizu-rivas_artificial_2024, lischka2023editorial, thasler2025journalistic}. Participation in AI-mediated journalism is thus fundamentally a group problem, unfolding across differently positioned actors whose roles, authority, and responsibilities are being actively reconfigured \cite{lischka2023editorial, thasler2025journalistic, van_drunen_editorial_2021}.

\paragraph{Participation under inter-organizational dependency.}
AI adoption in journalism is conditioned not only by newsroom dynamics but also by dependencies beyond the newsroom itself. AI adoption is uneven across organizations: large publishers are better able to build internal infrastructures and specialized teams, whereas smaller and local outlets are more often pushed toward commoditized tools, external funding schemes, and vendor-defined solutions \cite{quere_trust_2022, simon_uneasy_2022, simon_escape_2024}. These asymmetries generate lock-in effects and risks of infrastructural capture as platform companies extend their influence from distribution into core stages of news production \cite{fanta2020google, papaevangelou_funding_2024, simon_uneasy_2022}. Under such conditions, local participation does not necessarily translate into meaningful control over the most consequential layers of the AI stack \cite{corbett_power_2023, hansen_et_al_initial_2023, brogi2020monitoring, simon_uneasy_2022}. Research on large-scale scientific collaborations offers a useful analogue: actors from lesser-resourced institutions often faced structurally limited access to shared infrastructure and decision-making even when formally included \cite{luo2012human}. This pattern is particularly relevant for journalism, where smaller organizations may adopt AI as a survival strategy even when doing so requires partly relinquishing editorial control, compromising transparency, or accepting limited bargaining power over external providers \cite{paik_journalism_2025, quere_trust_2022, simon_artificial_2024, simon_escape_2024}. Accordingly, participatory outcomes must be evaluated not only within the newsroom, but also in relation to broader platform and vendor ecosystems \cite{drunen_safeguarding_2023, fanta2020google, papaevangelou_funding_2024, simon_uneasy_2022, simon_escape_2024}. In journalism, this interdependence has been conceptualized through participatory AI approaches attentive to ownership, control, and infrastructural dependency across micro, meso, and macro levels of the news ecosystem \cite{tseng_ownership_2025}. Building on this perspective, participation extends beyond internal newsroom inclusion toward ongoing coordination with external technology providers, where journalists contribute domain expertise, articulate professional constraints, and influence how tools evolve in practice. Under these conditions, participation becomes inherently cross-organizational, requiring alignment across editorial, technical, and commercial actors.

\paragraph{Participatory responses and their limits.}
%Indeed, 
Against these redistributions of authority, participatory approaches are often proposed as ways to reclaim agency, reduce knowledge silos, and align AI systems with journalistic values \cite{cooper_fitting_2024, corbett_power_2023, delgado_participatory_2023, tseng_ownership_2025}. Existing work argues that involving journalists and other affected actors in co-planning and co-design can support mutual learning, context-specific implementation, and collective ownership over newsroom change \cite{cooper_fitting_2024, feffer_preference_2023, fischer_understanding_2011, kelty_participatory_2017, tseng_ownership_2025}. Participatory approaches have also been framed as ways of creating shared spaces among journalists, developers, managers, and communities through bridge-building roles and structured collaboration around governance, evaluation, and oversight \cite{gerdes_participatory_2022, konya_democratic_2023, neghawi_linking_2023, xiao_it_2025}. At the same time, such efforts remain limited in practice \cite{delgado_participatory_2023, sloane_participation_2022, suresh_participation_2024}. In journalism, many proposals are still exploratory, unevenly institutionalized, and vulnerable to tokenistic or performative adoption \cite{hoque_towards_2024, tseng_ownership_2025, xiao_it_2025}. More critically, participatory AI scholarship warns of a \emph{participatory ceiling}, in which stakeholders may shape downstream implementation while remaining excluded from foundational decisions about adoption, optimization, and infrastructure \cite{suresh_participation_2024}. In such cases, participation may become an \emph{epistemic burden} rather than an empowering process \cite{birhane_power_2022}. In journalism, this ceiling is particularly visible in the misalignment between newsroom expectations and vendor-driven tool development: while journalists seek influence over editorial standards, alignment, verification practices, and accountability mechanisms, their participation is often restricted to feedback on surface-level functionality. This reinforces a structural gap between newsroom governance needs and the layers of the AI development that remain externally controlled.

\paragraph{Bridge to our empirical contribution.}
Taken together, the existing literature suggests that participation in AI-mediated journalism is shaped by newsroom hierarchies, cross-functional coordination, shifting divisions of labor, and unequal infrastructural dependencies \cite{cooper_fitting_2024, delgado_participatory_2023, suresh_participation_2024, tseng_ownership_2025, breed_social_1955, cools_news_2024, dodds2024impact, paik_journalism_2025, simon_uneasy_2022, thasler2025journalistic}. Participatory outcomes in journalism therefore depend not only on inclusion but on how participation is \emph{experienced} within newsrooms and organizational arrangements that unevenly distribute agency, accountability, and epistemic authority, and on whether system behavior, decisions, and accountability structures are sufficiently legible to support shared interpretation and contestation \cite{bratteteig_disentangling_2012, delgado_participatory_2023, suresh_participation_2024, dhanorkar2021needs, erickson2000social, holten2021can}.
%Together, 
These perspectives suggest that participatory AI in journalism must be understood as both layered (across normative, technical, and operational dimensions) and relational (unfolding through ongoing coordination between newsrooms, external technology providers, and the public) requiring frameworks that account for constrained influence within and beyond the newsroom.
 Building on these concerns empirically, our interviews show that newsroom stakeholders evaluate AI integration less through formal participation opportunities and more through the \emph{experienced} conditions of agency, accountability, and epistemic legibility. We introduce a construct for this empirical pattern in the Findings and design our framework to address it.

\section{Overview of Our Approach}
\label{sec:contributions}

This paper offers an empirical and conceptual account of participation in AI adoption in journalism. Drawing on interviews conducted in an independent local newsroom in continental Europe (\ref{sec:empirical-starting-point}), we show that stakeholders evaluate AI adoption not primarily through the formal availability of participation opportunities, but through the experienced conditions under which participation becomes meaningful: agency, epistemic legibility, and accountable governance. We describe this empirical pattern as a 'perception gap' between nominal inclusion and credible influence (\ref{subsec:perception-gap}).

Building on these findings, we develop the \emph{Gradual Voluntary Participation (GVP)} framework, in the form of a matrix (\ref{gvp-matrix}). The matrix conceptualizes participation across two dimensions: \emph{depth}, ranging from awareness to infrastructuring, and \emph{scope}, ranging from individual and team practices to organizational and public governance. Rather than treating participation as a single static ascending ladder, GVP frames it as uneven, role-dependent, and shaped by workload, authority, and organizational constraints.

We then derive \emph{five organizational principles} for making participation in AI adoption credible and sustainable in newsroom settings (\ref{gvp-principles}). These principles specify the conditions under which participation is more likely to be experienced as legitimate rather than burdensome, including recognition, resourcing, alignment, sovereignty, and accountability.

Taken together, the paper contributes: (1) an empirical account of how newsroom actors construe legitimate participation in AI adoption; (2) a two-dimensional framework for reasoning about participation across organizational depth and collective scope; and (3) a set of organizational principles for implementing participatory AI governance under real workplace constraints. The remainder of the paper presents the framework (\ref{gvp_framework}), illustrates its use through a newsroom scenario (\ref{subsec:sketching-application}), and discusses its limits, tensions, and implications for participatory AI governance under real newsroom constraints, including editorial accountability, time pressure, and dependence on external technology providers (\ref{subsec:challenges-limitations}).

\begin{figure}[!h]
    \centering
    \includegraphics[width=1\linewidth]{graph_contents.jpeg}
    \caption{Overview of paper contributions and their relationships. The three contributions build sequentially: empirical findings from newsroom interviews reveal the perception gap (Contribution 1), which motivates reconceptualizing participation as a navigable matrix rather than a unidirectional ladder (Contribution 2), which in turn motivates the derivation of five guiding principles for meaningful participation (Contribution 3). Together, these contributions inform the Gradual Voluntary Participation (GVP) framework for participatory AI integration in journalism described in Section \ref{gvp_framework}.}
    \label{fig:contents}
\end{figure}

\section{Contribution 1: Empirical Findings from a Local Independent Newsroom}
\label{sec:empirical-starting-point}

This paper starts from semi-structured interviews with journalists and editors about \emph{where} and \emph{how} they would want to participate in AI/LLM integration. Rather than assuming participation as a uniform normative ideal, we examine how it is articulated in practice through professional roles, organizational constraints, editorial responsibilities, and varying degrees of willingness to engage. Across the interviews, participation emerged not simply as a question of access to tools, but as a question of governance, accountability, knowledge, and recognition within newsroom work.
The framework developed in the following sections is derived analytically from these empirical patterns. We take these perspectives as the empirical problem-definition that the framework must satisfy: participation in AI adoption is heterogeneous, conditioned by workload and hierarchy, and experienced as legitimate only under specific organizational conditions. Importantly, participation and training were not the sole \emph{a priori} focus of the interviews: they emerged as salient concerns as interviewees reflected on how generative AI intersects with professional identity, editorial standards, accountability, and the future organization of newsroom work.

\subsection{Data and analytic approach}
We conducted 10 semi-structured, face-to-face interviews in an independent local newsroom in continental Europe, with journalists occupying different roles and the newsroom director. Participants included seven journalists, one journalist with an additional technical role, one newsroom director, and one social media content manager. The sample comprised six male and four female participants. Ages ranged from 30 to 60 years. Five participants were in the 50–60 age range, two were in the 40–49 range and three were in the 30–39 range (a mapping of the participants' demographics and roles can be found in Table \ref{tab:participants}).

\begin{table}[h]
\caption{Interview Participant Demographics and Roles}
\label{tab:participants}
\begin{tabular}{p{0.15\linewidth}p{0.1\linewidth}p{0.12\linewidth}p{0.55\linewidth}}
\toprule
Participant & Age & Gender & Role \\ 
\midrule
P1 & 50-60 & Male & Journalist (sport and local news) \\
P2 & 40-49 & Female & Journalist (sport) \\
P3 & 40-49 & Male & Tech editor and journalist \\
P4 & 30-39 & Female & Social media manager \\
P5 & 50-60 & Male & Editor and journalist \\
P6 & 40-49 & Female & Journalist (local and legal news) \\
P7 & 50-60 & Female & Newsroom director \\
P8 & 50-60 & Male & Journalist (local news)  \\
P9 & 50-60 & Male & Journalist (local news) \\
P10 & 30-39 & Male & Journalist (sport and local news) \\
\bottomrule
\end{tabular}
\end{table}

Each interview was conducted individually, in person, by two PhD researchers and lasted approximately one hour\footnote{All participants provided informed consent under ethics committee approval. Interviews were recorded, transcribed, anonymized, and then deleted. Institutional review board details are omitted for double-blind review and will be provided in the final version of the paper}. The interview protocol elicited preferences about (i) \emph{internal participation} (collaboration with newsroom management on policies, permissible uses, and rollout rules), (ii) \emph{knowledge-building participation} (internal/external training and peer learning to reduce knowledge silos), and (iii) \emph{external participation} (feedback loops with vendors or providers about tool limitations, professional needs, and workflow constraints). More broadly, interviews explored how newsroom professionals perceived and approached generative AI, including patterns of use, expectations, concerns, values, professional identity, editorial standards, trust, authorship, perceived alignment, and AI’s prospective role in journalism. Participation and training surfaced repeatedly within these discussions.

We analyzed the material through thematic synthesis focused on participation targets, willingness conditions, and resource constraints. Coding proceeded iteratively across transcripts, with themes refined through comparison across newsroom roles and positions.

We deliberately focused on a single newsroom community to retain a shared organizational context, meaning constraints, norms, workflows, and governance structures, while still capturing heterogeneity in roles, demographics, experience, and attitudes toward AI. This sampling strategy supports a contextually grounded analysis of how participation is understood and negotiated within a specific organizational environment and aligns with participatory design research emphasizing that meaningful design knowledge is situated \cite{suchman1987plans, dourish2001action, simonsen2014situated, ehn2017scandinavian}. We also selected a local independent newsroom because such settings remain comparatively under-examined in human-AI interaction and newsroom AI research \cite{usher2021news, napoli2017local}, despite often facing sharper resource constraints and distinct accountability dynamics than large national outlets \cite{abernathy2020will, radcliffe2017local}. Following qualitative design research, we prioritize \emph{transferability} over statistical generalizability \cite{lincoln1985naturalistic, flyvbjerg2006five}: thick description enables readers to judge relevance to their own settings, and intensive observation of critical cases can yield insights that broad surveys cannot \cite{stake1995case}. The framework is therefore grounded in this setting, while future work should test its transferability across multiple newsrooms, including larger organizations and different national contexts.

\subsection{An emergent finding: the "perception gap"}
\label{subsec:perception-gap}

Across interviews, journalists did not primarily evaluate generative AI adoption through model-centric criteria (e.g., output quality or efficiency), but through the \emph{experienced conditions} under which AI would enter newsroom practice.
In coding, we observed a recurring mismatch between how AI integration is often framed as a technical rollout and how participants assessed it through (i) perceived epistemic controllability, (ii) perceived procedural agency, and (iii) perceived legitimacy of participation demands.
We group this mismatch under the  \textbf{perception gap} theme: \emph{a divergence between technical deployment and journalists' perceived ability to understand, influence, and remain accountable for AI-mediated work}. 

\paragraph{Epistemic controllability: trust depends on perceived limits, provenance, and "red lines".}
Participants repeatedly emphasized that meaningful interaction requires profession-specific knowledge about what AI can and cannot be relied upon for, where it gets information from, and how to recognize failure modes.
Rather than requesting technical literacy for its own sake, journalists asked for normative and practical boundaries (e.g., "what I must not do with AI as a journalist"), as well as guidance on trust and tool selection%
\textbf{[P1, P2]}.

\paragraph{Procedural agency: participation was framed as influence over standards and accountability, not interface co-design.}
When participants expressed interest in participation, they emphasized contributing to internal policies and broader professional regulation, and some extended this to vendor-facing channels.
Notably, participation was repeatedly framed as normative work (standards, rules, accountability), explicitly distinguished from merely optimizing a product or service%
\textbf{[P4, P7, P2, P3]}.

\paragraph{Legitimacy of participation demands: willingness depends on recognition, time, and role boundaries.}
Even participants favorable to participation described it as conditional on workload protections and recognition as work (e.g., compensated time, explicit role credit), while others resisted any expectation of taking on coordinator-like responsibilities outside journalistic duties.
Time scarcity and asymmetric burdens made "participation" risk being perceived as unpaid extra labor rather than empowerment%
\textbf{[P6, P9, P10]}.

\paragraph{Why this matters for the framework.}
These patterns show that participation in AI adoption is not simply a matter of improving technical capability or offering formal opportunities for input. Rather, it is an organizational and group-work problem: journalists assess AI adoption through whether they have meaningful voice in decisions, whether systems are sufficiently legible to support professional judgment, and whether participation is recognized as part of newsroom work rather than added burden. The Gradual Voluntary Participation (GVP) framework is intended to address this problem by structuring participation across different levels of depth and scope and by making visible the organizational conditions under which participation becomes credible, coordinated, and sustainable.

\subsection{What "participation" meant to participants}
\label{subsec:participation-meaning}

Participants consistently framed participation less as "co-designing AI" in an abstract sense and more as institutional involvement in decisions that shape everyday work, especially rules, boundaries, and accountability arrangements.
Across interviews, participation clustered around three concrete themes:

\begin{enumerate}
    \item \textbf{Internal governance participation (newsroom-level):}
    contributing to policies and rules for AI use, defining acceptable use cases, and shaping implementation pathways with management.
    Participants referenced decisions about what AI uses should be permitted, disclosure requirements when AI is used in published work, verification standards for AI-generated content, and accountability escalation when AI outputs raise editorial concerns. Several emphasized wanting clarity on "what I must not do with AI as a journalist" and "what happens when something goes wrong" rather than technical configuration details. These forms of participation reflect journalism’s strong tradition of internal norm-setting around editorial standards, where legitimacy derives from collectively defined professional, beyond purely technical optimization.
    
    \item \textbf{Capacity-building participation (learning \& diffusion):}
    access to training and courses (internal and external), peer exchange, and mechanisms to prevent knowledge concentration in small groups.
    Participants distinguished between generic AI literacy courses (often offered by professional organizations like the Order of Journalists for continuing education credits\footnote{\url{https://www.odg.it/}}) and journalism-specific training covering practical use cases, regulatory/legal constraints, and professional standards. Many cited time scarcity as the primary barrier, expressing preference for internal newsroom training (integrated into work hours and "on-the-job" learning through experimentation) over external theoretical courses. Several noted existing knowledge silos where only certain staff members understand AI capabilities, creating uneven power dynamics.
    
    \item \textbf{External feedback participation (vendor-facing):}
    communicating newsroom needs, reporting limitations discovered in practice, and influencing tool evolution through feedback channels, ranging from bug reporting to structured consultations.
    Participants envisioned vendor-facing participation primarily as individual journalists providing feedback on tool limitations discovered through daily use, communicating professional constraints around sourcing verification, legal risk, editorial tone, and audience harm considerations. Some referenced structured consultation channels (pilot programs, user feedback sessions), while others imagined less formal bug reporting or feature requests. Several noted that tools often felt too generic or multi-purpose rather than tailored to journalism standards, creating frustration that could be addressed through direct professional input to developers. This external orientation reflects the ties between newsrooms and technology providers, where journalistic practice depends on tools developed outside the profession, making cross-organizational participation a necessary condition for maintaining editorial standards.
\end{enumerate}

\subsection{Patterns of willingness: participation as conditional, not binary}
\label{subsec:conditional-willingness}

Interviewees rarely expressed a simple "yes/no" stance toward participation.
Instead, they articulated \emph{conditional willingness}: participation was desirable under certain organizational conditions and undesirable under others.
We summarize these conditions as patterns that recur across the three arenas above.
The majority of participants (8 out of 10) explicitly framed willingness in conditional terms, using phrases like "I would like to participate, but..." or "it depends on whether...", rather than expressing unconditional enthusiasm or outright refusal.
Three patterns emerged:

\paragraph{Pattern A: Governance-first participation is widely desired, but with role-specific boundaries.}
Many participants expressed strong interest in contributing to internal rules and policies, often because these decisions affect accountability, autonomy, and workload distribution.
However, the preferred form of involvement varied by role and professional focus, and participants differed in how much time they could dedicate.
Field reporters emphasized needing clear guidelines more than involvement in creating them, while the newsroom director emphasized collaborative discussion without formal documentation. Several journalists wanted active roles in both internal standards and broader professional regulation, while others preferred consultative input without coordinator-level responsibility.

\emph{Illustrative paraphrase:}
%Journalists wanted a say in "what is allowed," "who decides," and "what happens when something goes wrong," but not necessarily in the technical configuration of the tool. One journalist stated wanting "a more active role in providing feedback to draft specific regulations for AI in journalism, both internally within my newsroom and more broadly for the profession" [P4]. Another emphasized wanting participation "to go beyond just optimizing/improving the product/service; it must also be normative/regulatory" [P7]. Conversely, one participant expressed interest in discussion but not wanting "an important and overly active role in this aspect" or coordinator responsibilities, framing participation as "an appendage to my job" as a journalist [P10].
Journalists wanted a say in "what is allowed," "who decides," and "what happens when something goes wrong," but not necessarily in the technical configuration of the tool. One participant working in social media management for the newsroom, who had engaged in multiple conversations with journalists about AI use, reported wanting "a more active role in giving feedback on regulations for AI in journalism, both here in our newsroom and for the profession more broadly. We need clear guidelines about what's acceptable and how we keep our standards when using these tools" [P4]. Another emphasized that participation must extend beyond product development: "For me, participation can't just be about the product or making it work better from a technical point of view. It has to be normative, regulatory. Journalists should be involved in setting the rules and must figure this out all together" [P7]. Conversely, one participant expressed interest in contributing while maintaining clear boundaries: "I'm interested in these discussions, but I don't want an overly active role in this, like being a coordinator or taking on extra responsibilities. I'm a journalist. Everything else is kind of an appendage to my actual job. I want to participate, but I need to protect my time for the work I was actually hired to do" [P10].

\emph{How this motivates our framework.}
The prominence of internal governance suggests that participation must be representable as policy and process involvement, not only tool co-design.
This motivates treating policy/process participation as a first-class participation mode in GVP, distinct from technical co-design, with varying depth levels that accommodate both active regulatory roles and consultative input, without mandatory coordination responsibilities.

\paragraph{Pattern B: Training is seen as participation infrastructure, but its value depends on access and institutional support.}
Participants described training courses as a mechanism to avoid knowledge silos and uneven power.
Interest was high when training was (i) integrated into working hours, (ii) recognized as part of the job rather than extracurricular effort, and (iii) connected to concrete newsroom policies (not generic AI literacy).
All 10 participants mentioned training or courses. Journalists across different experience levels wanted journalism-specific content covering both practical functionality and regulatory/legal constraints, distinguishing this from generic courses that "did not give me guidelines" and served primarily for earning professional credits [P1]. Several emphasized preferring practical, on-the-job experimentation ("trying things out with AI, understanding where it fails" [P7]) over theoretical instruction, and wanting internal newsroom training over external courses that were "hard to find time" for [P9].

\emph{Illustrative paraphrase:}
%Interviewees framed training as necessary to "speak the same language" and obtain meaningful participation across the newsroom, but resisted it when it became an additional burden on top of deadlines. One journalist wanted training "not so much about AI from a technical point of view, but about guidelines on what I must not do with AI as a journalist," noting that management needs to commit to training employees [P1]. Another emphasized wanting "specific courses on AI in journalism, including the regulatory/legal side" and needing "resources tailored to the specific profession" [P2]. However, time constraints were pervasive: "I would like them, but it's hard to find time; I never disconnect" [P9]. Knowledge silos were explicitly noted, with several participants describing uneven AI literacy creating barriers to collective participation.
Interviewees framed training as necessary to "speak the same language" and obtain meaningful participation across the newsroom, but resisted it when it became an additional burden on top of deadlines. One journalist explained their training needs: "I don't need training about AI from a technical point of view. What I need are clear guidelines about what I must not do with AI specifically as a journalist. And management needs to commit to training employees" [P1]. Another emphasized the need for profession-specific education: "I want specific courses on AI in journalism, not generic AI courses. I would like to learn about the regulatory and legal side, applied to our work and profession, specifically" [P2]. However, time constraints were pervasive: "I would like to do these courses, but it's very hard to find time. I'm already stretched thin, I never disconnect, I'm always on call" [P9]. Knowledge silos were explicitly noted, with several participants describing uneven AI literacy creating barriers to collective participation.

\emph{How this motivates our framework.}
Training functions as participation scaffolding: it enables meaningful engagement with governance and tool use, and it shapes who can participate.
This motivates GVP's contextual knowledge scaffolding principle, emphasizing that training must be integrated into work time, tailored to professional contexts (not generic AI literacy), and provided through multiple modalities (courses, peer learning, on-the-job experimentation) to accommodate diverse learning preferences and time constraints.

\paragraph{Pattern C: Vendor-facing participation is polarizing and reveals workload/recognition conditions.}
Responses diverged most strongly on direct participation with tech companies and providers. Some interviewees welcomed the opportunity to influence tools using professional expertise; others saw vendor feedback as extra labor that should be compensated or formally recognized. Several framed vendor-facing participation as only feasible if the organization created protected time and clear incentives. Divergence appeared linked to workload intensity, prior experiences with vendors not meeting professional standards, and frustration with tools being too generic rather than journalism-specific. Those most enthusiastic about vendor participation emphasized normative/regulatory influence, not just product optimization.

\emph{Illustrative paraphrase (pro):}
%Participants valued the chance to communicate professional constraints and newsroom standards directly to tool-makers. One journalist expressed "strong interest in having a more participatory role" both internally and "having a more direct line with producers/vendors/tech companies to communicate my feedback on the tools and my needs as a professional" [P2]. Another emphasized that participation should address regulatory concerns: "it's interesting when companies propose that their employees and sector experts actively participate in the development/building of the tools they will then use" [P7]. Professional constraints emphasized included sourcing verification requirements, legal risk boundaries, editorial tone standards, and preventing audience harm.
Some participants valued the chance to communicate professional constraints and newsroom standards directly to tool-makers, particularly those with direct technical experience. One journalist with hands-on experience using both LLMs and AI-based image and voice generation tools explained: "I've been experimenting quite a bit with these technologies and I think I understand how they work in general terms and I was able to notice some limitations as well. I'd really like to be involved in discussions with tech companies to provide direct feedback about what actually happens when you try to use these tools for journalism work" [P3]. One journalist expressed enthusiasm for direct engagement: "I have strong interest in having a more participatory role, both internally in how we use and regulate these systems in the newsroom, and in having a more direct line with producers and tech companies to communicate my feedback on the tools and my needs as a professional. About errors I may find as well" [P2]. Another emphasized that participation should extend to regulatory dimensions: "It is really interesting when companies propose that their employees actively participate in the development and building of the tools they will then use, thinking about the professional standards and constraints we need from the start" [P7]. Professional constraints emphasized included sourcing verification requirements, legal risk boundaries, editorial tone standards, and preventing audience harm.

\emph{Illustrative paraphrase (conditional):}
Others were willing only if it counted as work: paid time, explicit role credit, or workload reductions elsewhere. One journalist stated: "it would also be interesting to participate, to have a more active role, but it is extra work and should be compensated/considered working time. I would not want it to be imposed on me as another job" [P6]. Recognition forms proposed included formal working time allocation, compensation for feedback contributions, and acknowledgment in role descriptions or performance evaluations.

\emph{Illustrative paraphrase (resistant):}
Some rejected it as "not my job," preferring vendor interaction to be delegated to dedicated roles. One participant stated: "we have to consider that I do another job; I wouldn't want it imposed. I am a journalist, everything else I consider an appendage to my job, so I would not want to have an important and overly active role in this aspect" [P10]. Suggested delegation targets included product managers, innovation teams, newsroom standards leads, or professional organization representatives who could aggregate and communicate feedback on behalf of multiple journalists.

\emph{How this motivates our framework.}
This divergence highlights the need to model participation as voluntary and resource-sensitive, with explicit handling of recognition, compensation, and role boundaries.
This motivates GVP's recognition-without-coercion principle and its emphasis on voluntary participation with transparent time/resource requirements. The framework must accommodate both individual direct engagement and mediated participation through dedicated roles or professional organizations, allowing journalists to opt into vendor-facing participation only when adequately supported and recognized, rather than imposing it as a default expectation.
\section{Contribution 2: The Design Space - Rethinking Participation in Journalism, from Ladder to Matrix}
\label{gvp-matrix}
%For the aforementioned reasons, evaluating participatory AI design requires more than procedural inclusion, demanding attention to how the perception of secondary stakeholders changes at the intersection of voluntary and gradual participation. 

\subsection{Beyond the Ladder: Depth and Scope}

Arnstein's influential ladder of participation \cite{arnstein_ladder_1969} conceptualizes participation as a vertical progression from non-participation, through tokenism, to citizen power. While foundational, this model assumes a largely unidirectional movement toward increased power and does not readily account for the heterogeneous, overlapping, and role-dependent forms of participation that characterize AI adoption in organizational settings. In newsrooms, participation is not simply a matter of moving upward along a single scale, but is instead shaped by professional hierarchy, workload, domain expertise, and the distribution of decision-making authority across actors within and beyond the organization. As a result, ladder-based metaphors are limited in their ability to capture how participation is organized, constrained, and negotiated across multiple dimensions of collective work.

Other frameworks, such as the one proposed by Suresh et al. \cite{suresh_participation_2024}, tackle the need to conceptualize "participation in the age of foundation models", hence visualizing participation as potentially happening at different institutional and design layers on top of the chosen "foundation" model, adding a sense of \textit{depth} rather than only evolving gradually like across the rungs of a ladder. 
Similarly, in the context of journalism, a mapping of the evolving dynamics across stakeholders dealing with AI implementation was proposed by Tseng et al. \cite{tseng_ownership_2025}, who highlight different \textit{scopes} of participation at "micro", "meso", and "macro" levels for newsrooms. 

We advocate for PD design frameworks that are built on top of these concepts across two main dimensions of a participatory matrix:

\begin{itemize}
    \item \textbf{Depth:} From awareness and consultation, through procedures configuration, to co-creation and infrastructuring.
    \item \textbf{Scope:} From individual practices, through organizational policies, to sectoral and public standards.
\end{itemize}

Stakeholders do not necessarily move through these dimensions in a linear way. A journalist might deeply engage with prompt engineering (high depth, narrow scope) while minimally participating in industry-wide policy discussions (low depth, broad scope). This variation creates natural "experimental" conditions for understanding how different participation configurations influence perception. The journalist with deep technical engagement but minimal policy involvement likely perceives AI differently than another one with inverse participation patterns. These differences become analyzable, rather than problematic, when the participation framework itself supports such multidirectional movement as an expected design feature, instead of treating it as deviance from a prescribed path.
Unlike hierarchical models that assume unidirectional progression, this matrix creates a navigable participation space: stakeholders can explore multiple pathways simultaneously, moving vertically (increasing depth within current scope), horizontally (expanding scope at current depth), or diagonally (combined shifts across both dimensions). This multidirectional movement reflects how participation unfolds in practice, as actors engage across different roles, decision arenas, and levels of influence rather than following a single prescribed sequence. 
The matrix thus functions as both a conceptual model and an organizational tool: it makes visible where participation is located, what forms of involvement are available under different conditions, and how participation is distributed across actors, roles, and arenas of coordination, making peer participation patterns visible across the landscape.

Prior work suggests that structured spatial representations can help users understand complex multidimensional relationships more effectively than fragmented alternatives \cite{reeder2008expandable}. Relatedly, feedforward mechanisms can support exploration by previewing possible consequences before commitment, especially in open-ended tasks where actors may consider multiple participation pathways \cite{djajadiningrat2002but, vermeulen2013crossing, terry2002}.

The proposed matrix structure responds to what Bratteteig and Verne \cite{bratteteig_does_2018} identify as AI's core challenge to traditional PD: systems that change continuously through learning require participation models that accommodate ongoing, multi-directional engagement rather than fixed, unidirectional progression.

\subsection{The Voluntary Principle and Self-Determination Theory}

Mandatory participation creates what Birhane et al. call "epistemic burden" \cite{birhane_power_2022}, requiring stakeholders to invest cognitive and temporal resources without compensation. In practice, participation in AI design, adoption and governance is often presented as voluntary while remaining structurally compulsory through organizational expectations, role responsibilities, and the coordination demands of collective work. Non-voluntary participation may take the form of mandatory tool use without meaningful avenues for feedback, implicit expectations to engage in consultation processes as part of professional duties, or alignment activities that are formally voluntary but tied to performance evaluation, workload, or organizational compliance \cite{groves_going_2023}. An alternative is to center participation on voluntary choice while still recognizing contributions, thereby avoiding coercion while maintaining engagement.

This approach draws from self-determination theory, which identifies three basic psychological needs driving intrinsic motivation: autonomy (the need to volitionally endorse one's actions), competence (the need to feel effective), and relatedness (the need to connect with others) \cite{deci2012self}. By preserving choice while providing recognition, voluntary participation sustains engagement without exhausting stakeholders. The voluntary principle suggests that chosen participation produces fundamentally different perceptions than mandated engagement, a distinction critical for understanding why similar participation activities produce varying perceptual outcomes across different organizational contexts.

%% FIVE MECHANISMS

\section{Contribution 3: Five Principles for Gradual and Voluntary Participation in Newsrooms}
\label{gvp-principles}
To center participants' lived experience within a framework for AI governance, a set of organizational principles is needed to ensure that even gradual and voluntary participation remains meaningful within collective work, rather than tokenistic or performative. In this section, we present the principles as well as corresponding operational steps (concrete recommendations) that help translate theory into practice. A detailed mapping of how each principle is grounded in the empirical findings is provided in Appendix~\ref{supplementary-material-mapping-gvp}, Table~\ref{tab:gvp_mapping}. 

\subsection{Recognition Without Coercion}

We call the first interaction design principle Recognition Without Coercion, reflecting what Delgado et al.'s participation spectrum suggests \cite{delgado_participatory_2023}: multiple valid approaches rather than hierarchical progression. Transparent certification systems can document participation levels, providing visible feedback about engagement status without creating prerequisites.

\textbf{Concrete Recommendations:}
\begin{itemize}
\item Develop transparent and shared certification systems and interfaces that document participation levels without creating professional prerequisites.
\item Establish multiple navigable pathways to professional development, with AI engagement as an option, not an obligation.
\item Conduct stakeholder mapping and power analysis to surface hidden asymmetries and contextualize recognition processes.
\end{itemize}

We hypothesize that when journalists observe respected colleagues engaging in value alignment and other human-supervised phases of the AI-augmented workflow, these actions could foster trust even among non-participants. 

\subsection{Structures for Continuous Value Alignment}

Continuous Value Alignment aims to show system responsiveness through regular value collection cycles and threshold-triggered policy reviews, where different stakeholders can engage directly through interactive touchpoints to propose values, and have their opinions collected and visualized by the system. Drawing on participatory action research \cite{hayes_knowing_2014, rasmussen_action_2004}, values must emerge from ongoing negotiation among affected parties through responsive interaction mechanisms rather than static imposition.

\textbf{Concrete Recommendations:}
\begin{itemize}
\item Implement regular value collection cycles capturing different stakeholder priorities.
\item Create threshold feedback mechanisms triggering policy reviews when values shift significantly.
\item Enable bottom-up value proposition through accessible input affordances, while maintaining institutional coherence.
\item Incorporate value collection logs to reveal implicit tensions and inform alignment cycles.
\end{itemize}

Supporting alignment also requires mechanisms that make the effects of participation traceable over time. Prior work on history flow visualizations \cite{viegas2004} and knowledge transfer graphs \cite{zhao2018gender} shows how contributions can be followed as they propagate through collaborative processes, directly applicable to showing journalists how their value propositions influence policy evolution, workflows, or governance decisions.

We hypothesize that this ongoing adaptation could influence perception by revealing AI as negotiated rather than imposed. Documenting value evolution creates consciousness, allowing journalists and readers to trace how collective priorities shaped the system’s use and implementation.

\subsection{Contextual Knowledge Scaffolding}

Contextual Knowledge Scaffolding addresses information asymmetries through modular learning resources linked to specific interaction and decision points in the AI-augmented content generation pipeline. Drawing on situated learning theory \cite{Lave_Wenger_1991}, this interaction principle recognizes that perception emerges from legitimate peripheral participation, with knowledge acquisition occurring through graduated involvement rather than front-loaded prerequisites.

\textbf{Concrete Recommendations:}
\begin{itemize}
\item Develop modular knowledge resources linked to specific interaction and decision points.
\item Create multidirectional transparency where all stakeholders can access the same learning information.
\item Design peer-learning networks supplementing formal training.
\item Integrate scaffolding activities and interfaces with stakeholder mapping to ensure accessibility across differing expertise levels.
\end{itemize}

This scaffolding approach builds on Yildirim et al.'s finding that providing capability information at ideation-time helps stakeholders generate a broader range of concepts \cite{yildirim2023creating}, though ensuring technical feasibility requires active collaboration with data scientists. In journalism contexts, just-in-time learning 
resources linked to specific decision points (e.g., when selecting AI summarization 
parameters) could reduce epistemic burden while enabling informed participation.

Empirical work supports this just-in-time approach to knowledge support. Grossman and Fitzmaurice found that contextual video assistance helped users complete substantially more unfamiliar tasks than traditional online help, with better retention over time \cite{grossman2010}. Similarly, Chilana et al.'s \emph{LemonAid} system showed that contextual, selection-based help could surface relevant guidance for many user queries \cite{chilana2012}. These findings reinforce GVP's emphasis on embedding knowledge support within ongoing work practices, rather than treating competence as something that must be fully established in advance.

We hypothesize that multidirectional transparency will mitigate the mystification that fosters epistemic capture. Furthermore, a progressive introduction of complexity will enable diverse stakeholders (e.g., journalists and readers) to develop nuanced mental models at their own pace. 

\subsection{Sovereignty Preservation}

Sovereignty Preservation maintains clear interaction hierarchies between human judgment and AI output through unambiguous override mechanisms. Clear interface hierarchies, where human judgment supersedes algorithmic output, contribute to human agency preservation.

\textbf{Concrete Recommendations:}
\begin{itemize}
\item Establish unambiguous override affordances, mechanisms and standards at every interaction and decision point.
\item Create documentation systems recording when and why human judgment diverges from AI suggestions.
\item Build accountability structures, such as certification seals visible in user interfaces, ensuring humans remain responsible for outcomes.
\item Adopt graduated implementation pathways that begin with voluntary pilots and expand incrementally based on demonstrated trust and value.
\end{itemize}

These override mechanisms align with foundational human-AI interaction principles. Horvitz  \cite{horvitz1999principles} and later Amershi et al. \cite{amershi2019} proposed that systems must support efficient invocation, termination, dismissal, and correction of AI services, minimizing costs when users reject automated outputs. Buçinca et al. further demonstrated that cognitive forcing functions, which require analytical engagement before revealing AI recommendations, reduce overreliance, and support appropriate human authority \cite{Bucinca2021}.

These patterns also align with Berge et al.'s findings on human-AI collaboration in professional settings \cite{berge2023designing}: preserving professional autonomy in high-stakes domains requires workflows that sustain human judgment throughout ongoing work, rather than relying on post-hoc override options that arrive too late in the workflow.

We hypothesize that documenting instances where professionals diverge from AI recommendations will provide tangible evidence of professional autonomy. Making this autonomy visible may shape stakeholders’ perceptions by clarifying that AI outputs are advisory rather than mandatory. 

\subsection{Transparent Accountability}

Transparent Accountability requires more than reporting outcomes; it involves making the processes behind those outcomes visible and explorable to different stakeholders through interactive mechanisms and
interfaces. This can be achieved through multi-level disclosure tailored to the needs of each group, creating what the algorithmic accountability literature \cite{leppanen2017data} demands: process visibility and legibility through interaction beyond outcome transparency.

\textbf{Concrete Recommendations:}
\begin{itemize}

\item Create interactive audit trails that document how inputs shape outputs.
\item Establish clear attribution linking decisions to responsible parties.
\item Develop accessible documentation and interfaces translating technical details into understandable terms.
\end{itemize}

This multi-level approach operationalizes Dhanorkar et al.'s framework \cite{dhanorkar2021needs}, which originally demonstrated that data scientists, product managers, and business owners require different explanations at various AI lifecycle stages, to the newsroom context. This application recognizes that developers, editors, journalists, and readers similarly possess distinct transparency needs, confirming that a single explanation mechanism cannot serve all stakeholder groups simultaneously.

We hypothesize that such audit trails would allow journalists, for example, to trace how their contributions influence the system’s behavior. This traceability would shift perceptions from speculation to evidence: rather than guessing whether participation matters, stakeholders would directly observe its effects on AI behavior. 

\section{GVP: The Gradual Voluntary Participation Framework}
\label{gvp_framework}

Building on these contributions, we propose a framework to navigate stakeholders' perception of participatory AI across different levels of voluntary and gradual participation, the "Gradual Voluntary Participation (GVP)" framework. We propose a checkerboard-like matrix structure in which the five organizational principles apply across different levels of depth (X axis) and scope (Y axis). This representation draws on prior work showing that structured spatial arrangements can support understanding of complex multidimensional relationships \cite{reeder2008expandable} and that feedforward can help actors anticipate possible consequences before commitment \cite{djajadiningrat2002but}. Framed this way, the matrix supports what Niemantsverdriet et al. \cite{niemantsverdriet2019} call "awareness in shared systems" by making participation, interdependence, and accountability more visible across roles and sites of coordination.

%To unfold and navigate the framework's chessboard, we chose journalism as the scenario. In this analogy that functions as a visual interaction model, stakeholders are represented by chess pieces: rooks as journalists, knights as tech developers, kings as editors and decision makers, queens as external organizations such as professional associations or academic institutions,  pawns as readers and public representatives

%We adopt the chessboard as a visual, design-motivated metaphor not only for its visual and conceptual clarity but also because, in chess, pieces move in multiple directions (vertically, horizontally, diagonally, or in an "L" shape), interaction patterns that mirror the flexible, nonlinear ways in which stakeholders can move across participation dimensions within our design framework. 
We adopt the checkerboard as a visual metaphor for its capacity to render participation legible as a spatial configuration, making the framework visually concrete much as the ladder did for earlier models of participation. The board supports orientation by making positions, trajectories, and co-existing forms of engagement visible at a glance, enabling stakeholders to situate themselves relative to others without prescribing sequences or outcomes. 
In this sense, the checkerboard interaction model emphasizes stakeholder agency in choosing pathways, discovering alternatives, and coordinating with peers across the participation landscape, beyond simply "climbing a ladder" of participation.

\subsection{Navigating the Checkerboard Matrix}
\label{subsec:navigating-chessboard}

\begin{figure} [!h]
    \centering
    \includegraphics[width=1\linewidth]{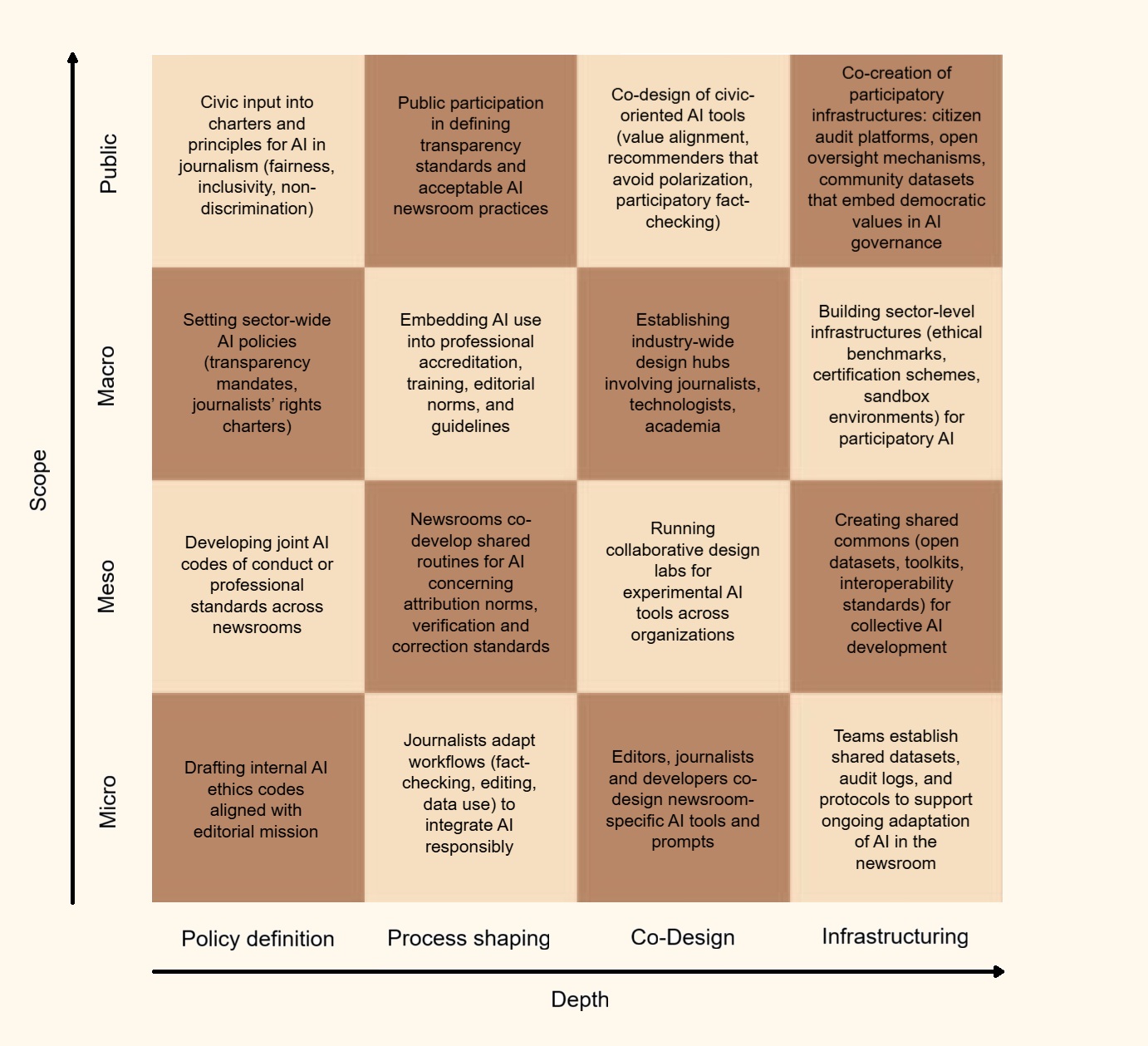}
    \caption{The GVP checkerboard, where each quadrant represents a distinct degree of participation across various scopes and depths of AI integration in journalism. }
    \label{fig:chessboard}
\end{figure}

Imagine a scenario in which a newsroom decides to engage in a participatory process to implement AI in journalistic workflows. This can take different forms, depending on the needs, values, resources, and interests of the participants. In particular, individual journalists, as the most directly influenced stakeholders, may feel the need to participate (or not) at different stages of the process, depending on their background and willingness to work with other actors (such as editors, developers, or external organizations). GVP's proposal is to empower affected stakeholders, in this case journalists, to do so at their own will and pace, across various depths and scopes. For this to positively influence their final perception towards the participation process result, it is necessary to ensure that the five fundamental principles described above are respected while moving across the checkerboard (see Figure \ref{fig:chessboard}). 

The X axis of the checkerboard matrix illustrates the various depth levels of possible participation, progressively moving from preliminary and planning practices to practical co-design and case-specific infrastructuring activities. These are:  \textit{policy definition, process shaping, co-design,} and \textit{infrastructuring}, mirroring Suresh et al.'s blueprint for "participation in the age of foundation models" \cite{suresh_participation_2024}.
The Y axis defines the scope of the participation, as it can happen in well-defined and confined scenarios, such as individual newsrooms, as well as expand into shared and public domains, encompassing various levels of the journalism industry. This axis is defined building over Tseng et al.'s description of tensions in journalism across \textit{micro, meso}, and \textit{macro} levels \cite{tseng_ownership_2025}, with the addition of a \textit{public }level, oriented at the collaboration with readers. 

To clarify how these two dimensions can be interpreted in practice, Appendix~\ref{appendix:gvp_operationalization}, Table~\ref{tab:gvp_operationalization}, provides an operationalization of each depth and scope level through working definitions, observable indicators, and example governance artifacts.
%% METHODOLOGICAL FRAMEWORK

\subsection{Sketching a possible GVP application}
\label{subsec:sketching-application}

The following scenario illustrates GVP's interaction patterns in practice, showing how stakeholders might engage with participation through specific interface touchpoints and feedback mechanisms. 
We argue that when a hypothetical newsroom adopts the Gradual Voluntary Participation (GVP) framework to guide their AI  governance and implementation processes, participation unfolds as a fluid, multi-level process that evolves voluntarily rather than through prescription. 
We present here just one of the possible examples of this evolution; for reasons of space, we offer only a brief sketch, leaving deeper scenario exploration for future work. 

At the micro level, in a small local newsroom participation may begin with a few journalists who voluntarily experiment with an AI-driven summarization tool.
Participation is recognized but never imposed, reinforcing the first GVP rule of recognition without coercion \textit{(Principle 1)}, and immediately shaping perceptions of AI as an opportunity rather than an imposition. Through regular and voluntary alignment meetings, staff members discuss emerging tensions such as accuracy, tone, or sourcing, adjusting both workflows and values in line with the second principle of continuous alignment \textit{(Principle 2)}. Internal guides and short workshops scaffold knowledge, allowing participants to understand system behavior at their own pace and depth \textit{(Principle 3)}. Crucially, journalists retain full editorial authority, preserving professional sovereignty \textit{(Principle 4)}, while transparent records of edits and overrides ensure traceability \textit{(Principle 5)}. These simple design choices, mirroring real-world initiatives by the Associated Press \footnote{\url{https://www.ap.org/the-definitive-source/behind-the-news/standards-around-generative-ai/} }, The Financial Times\footnote{\url{https://www.ft.com/content/18337836-7c5f-42bd-a57a-24cdbd06ec51}}, the BBC\footnote{\url{https://www.broadcastnow.co.uk/production-and-post/bbc-sets-protocol-for-generative-ai-content/5200816.article}}, or The New York Times\footnote{\url{https://www.theverge.com/news/613989/new-york-times-internal-ai-tools-echo}} turn experimentation into a safe, visible form of agency. 

As experiences and resources mature, collaboration can expand to the meso level, where several newsrooms can join forces to co-design shared routines, ethical standards, and open repositories of annotated datasets or prompt templates. Every contribution is credited in public documentation, reinforcing recognition and mutual trust, while periodic consortium meetings review earlier decisions to sustain value alignment. Shared repositories and modular training resources act as living scaffolds for newcomers, who can understand past deliberations without losing their own editorial independence, ensuring sovereignty preservation and transparent accountability. This collective experimentation, seen in collaborations such as JournalismAI\footnote{\url{https://www.lse.ac.uk/media-and-communications/polis/JournalismAI}} or the Partnership on AI\footnote{\url{https://partnershiponai.org/ai-for-newsrooms/}}, strengthens inter-organizational trust and transforms isolated trials into cumulative learning. In this context, Tseng et al. propose the creation of a conceptual News Tooling Alliance (NTA) \cite{tseng_ownership_2025}, to empower newsrooms to retain agency and ownership through the participatory process. 

A useful contrast to this trajectory is offered by recent conflicts around the use of journalistic content in foundation model training. As illustrated by Suresh et al., individual actors such as news organizations or journalists often lack the leverage to influence data practices at the foundation layer when acting alone, as exemplified by lawsuits like the one filed by The New York Times against OpenAI and Microsoft \cite{suresh_participation_2024}. In the absence of coordinated structures, such efforts risk remaining fragmented, costly, and ultimately ineffective in shifting industry norms. From a GVP perspective, this scenario highlights what happens when participation does not scale beyond isolated micro or meso engagements.

In contrast, the expansion toward macro-level participation envisioned in GVP creates the conditions for collective stakeholder power: newsrooms, unions, and professional organizations can collaboratively define data governance standards, negotiate licensing agreements, or even contribute to shared, curated datasets under explicit terms of use. This aligns with the “subfloor” dynamics described by Suresh et al., where coordinated group participation enables stakeholders to exert influence on the foundation layer itself \cite{suresh_participation_2024}. In this sense, GVP reframes participation as a proactive, infrastructural process through which journalistic actors can shape the conditions of both AI development and integration.

From there, participation in GVP naturally extends to the macro and public quadrants of the board, where a wider range of stakeholders such as journalists, readers, researchers, and civic groups can engage in open dialogues about AI’s role in news production. Readers may give feedback on AI-assisted articles or be consulted for value alignment, while visible provenance markers and open audit trails (as promoted by initiatives like C2PA\footnote{\url{https://c2pa.org/}} or Project Origin\footnote{\url{https://www.originproject.info/}}) make the process intelligible to audiences. Editorial sovereignty remains intact, yet the transparent sharing of decisions and rationales improves accountability and trust, potentially gradually shifting public perception from skepticism to informed participation. 
At this level, we argue that the participatory effort resembles what Bratteteig et al. describe as \textit{infrastructuring} \cite{bratteteig_what_2016}, a sustained, collective process where journalists, organizations, and readers co-evolve practices through voluntary, aligned, and transparent engagement. Across these moves throughout the checkerboard, the back-and-forth exchange of journalists refining tools, organizations collecting insights, and readers evaluating outcomes, is at the core of GVP: such dynamics can potentially transform resistance into curiosity and position AI as a collectively and voluntarily governed innovation built on gradual and shared legitimacy, influencing perception.

\subsection{Challenges and Limitations}
\label{subsec:challenges-limitations}

Adopting a framework like GVP, while proactive and user-oriented, introduces several practical and conceptual challenges. A primary difficulty lies in operationalizing the five core principles across different depth–scope contexts, which may encounter resistance depending on where and how stakeholders choose to engage. For example, recurrent alignment mechanisms may be difficult to sustain in resource-constrained settings, while contextual knowledge scaffolding could create tensions between participants with differing expertise or priorities.
A further tension emerges between GVP’s emphasis on gradual, voluntary participation on one hand and organizational pressures for rapid AI adoption on the other. Newsrooms and other institutions often face external urgency driven by competitive dynamics, technological change, or managerial timelines, which may conflict with the slower, exploratory rhythms required for reflective participation. In such contexts, gradualism may be perceived by some stakeholders as delaying action rather than enabling legitimacy, particularly when participation is framed as optional rather than mandatory. This creates a risk that GVP is either bypassed in favor of fast deployment or selectively applied in ways that undermine its core principles. These pressures may also blur the boundary between voluntary and non-voluntary participation. For instance, stakeholders may be formally invited to engage while lacking a meaningful option to decline, or participation may be implicitly required to ensure compliance with new AI systems. In such cases, participation risks becoming performative or coercive, undermining the perceptual and legitimacy-building goals that GVP seeks to support.
The dynamic, non-linear nature of participation trajectories also presents challenges. Stakeholders may instantiate unexpected movement patterns, contest roles, or experience conflicts when voluntary engagement intersects with organizational obligations. 
Finally, evaluating perception remains a persistent challenge \cite{bossen2010, bratteteig_what_2016}. As GVP emphasizes voluntary and gradual engagement rather than prescriptive outcomes, traditional metrics of "success" or completion may be inadequate. Observed shifts in understanding, negotiation, and alignment can be subtle, uneven, or temporary, further complicating assessment.
A further conceptual limitation concerns the meaning of governance itself. GVP is designed to support gradual, voluntary, and perception-sensitive participation, but it does not in itself specify enforcement mechanisms or consequences for actors who disregard agreed processes or norms. In this sense, GVP may be better understood as a governance-support framework useful to understand, plan and navigate participatory AI practices in multi-stakeholder domains such as journalism, rather than a fully institutionalized governance regime. This creates a risk that, in the absence of accountability structures, participation remains weak, symbolic, or performative rather than substantively shaping organizational AI practices. Acknowledging this limitation is important, as voluntary engagement alone does not guarantee compliance, responsibility, or institutional follow-through.
Overall, while GVP offers mechanisms for shaping positive perceptions and supporting reflective engagement, these same features can introduce unpredictability and context dependence. Competitive, high-pressure, or resource-limited environments may amplify tensions, potentially limiting the framework’s effectiveness. Careful facilitation, contextual adaptation, and iterative reflection are therefore essential to mitigate these challenges without undermining GVP’s core principles.

\section{Conclusions}

Building on the empirical identification of a \textit{perception gap} between nominal inclusion and credible influence, this paper introduced the Gradual Voluntary Participation (GVP) framework as a way of organizing participation across \textit{depth} and \textit{scope}, and articulated five principles for making such participation credible in newsroom AI adoption. Taken together, these contributions frame participation in AI adoption  in journalism not as a matter of tool acceptance alone, but as an organizational process shaped by gradualism, voluntariness, and the uneven distribution of voice, legitimacy, and influence. In line with PD evaluation debates that emphasize situated user gains \cite{bossen2010} over universal metrics, GVP treats negotiated participation and shifts in how participation is experienced as meaningful outcomes in their own right.

From this perspective, participation in AI should be understood as a multidimensional organizational phenomenon rather than a binary of trust versus distrust. By making layers of influence visible, recognizing voluntary engagement, and documenting outcomes, GVP reframes familiar participatory limitations such as ceilings, burdens, and performative consultation as conditions that must be explicitly addressed in organizing AI adoption. Grounded in self-determination theory, it further suggests that autonomy-supportive participation may foster more constructive orientations toward AI than mandated engagement. GVP also extends discussion of participation in complex organizational settings by treating AI governance as a problem of coordination across differently positioned actors rather than only dyadic human-AI interaction. In doing so, it brings together concerns such as navigation \cite{reeder2008expandable}, provenance \cite{viegas2004}, awareness \cite{erickson2000social}, scaffolding \cite{chilana2012}, and override \cite{amershi2019} within a single framework for reasoning about participation where professional roles, situated expertise, and decision-making authority are unevenly distributed, extending feedforward principles \cite{djajadiningrat2002but} from tool use to collective participation.

Methodologically, we derive a conceptual framework from qualitative data rather than presenting a validated intervention \cite{flyvbjerg2006five, lincoln1985naturalistic}. As in prior work that developed conceptual frameworks from situated qualitative inquiry \cite{lustig2024, spiel2021, chung2022}, we prioritize transferability through thick description over statistical generalizability, treating GVP as an analytically grounded account of participation under specific organizational conditions.

Future work should extend this contribution in two directions. First, empirical \textit{in-situ} studies in newsroom settings can examine how gradual and voluntary participation shapes journalists' perceptions, experiences of AI, and longer-term adoption practices. Second, while developed for journalism, GVP may provide a useful basis for studying participatory AI governance in other professional domains adopting large language models, especially where influence is distributed across multiple actors, levels of work, and institutional constraints.

\section* {Acknowledgments}
This work was funded by the EU Horizon Europe program, Marie Skłodowska-Curie Actions (MSCA), alignAI project (grant  101169473).
We thank our colleagues at the Social Computing Group for their feedback and support throughout this work.

\section* {Generative AI Disclosure Statement}
In the preparation of this manuscript, large language models were exclusively used for formatting assistance and language support, including designing images, refining paragraph layout and spacing, and performing proofreading, spelling, grammar, and lexical checks. All substantive content, analysis, and interpretations were conceived and written by the authors.

\bibliographystyle{ACM-Reference-Format}
\bibliography{bibliography}

\clearpage

\appendix
\section{Supplementary Material: Mapping of Empirical Findings to the Five GVP Principles}
\label{supplementary-material-mapping-gvp}

In this appendix, Table~\ref{tab:gvp_mapping} maps the five GVP principles to their empirical grounding in the interview findings. The table identifies the \emph{dominant} empirical grounding of each principle across perception-gap dimensions and participation patterns. For each principle, it indicates the most relevant perception-gap dimensions (\ref{subsec:perception-gap}), participation patterns (\ref{subsec:conditional-willingness}), and participant statements, and then makes explicit the analytical move through which these situated findings are elaborated into design-level commitments. The table should therefore be read not as a mechanical derivation from data to framework, but as a structured account of how recurring empirical concerns informed the conceptual development of GVP.

\begin{longtable}{p{0.15\linewidth}p{0.16\linewidth}p{0.34\linewidth}p{0.35\linewidth}}
\caption{Mapping of empirical findings to the five GVP principles. The table identifies the dominant empirical grounding of each principle, while also indicating overlap across perception-gap dimensions and the analytical move from situated findings to design-level commitments.}\label{tab:gvp_mapping}\\
\toprule
\textbf{Principle} & \textbf{Dominant empirical grounding} & \textbf{Empirical grounding in the interviews} & \textbf{Framework elaboration}\\
\midrule
\endhead

P1: Recognition Without Coercion
&
Primarily grounded in the \emph{legitimacy of participation demands} dimension of the perception gap (\ref{subsec:perception-gap}), especially Pattern~C, with overlap with Pattern~A.
&
Participants consistently framed willingness to participate as conditional on workload protections, recognition as work, and respect for role boundaries. P6 stated that vendor-facing participation "is extra work and should be compensated/considered working time. I would not want it to be imposed on me as another job". P10 articulated the role-boundary issue even more sharply: "I am a journalist, everything else I consider an appendage to my job, so I would not want to have an important and overly active role in this aspect". Pattern~A reinforces this same concern in the governance arena: P10 expressed interest in participating in discussions while insisting on protecting time "for the work I was actually hired to do". P9 similarly described pervasive time scarcity as a barrier across participation venues. The external feedback participation dimension (\ref{subsec:participation-meaning}) provided the most detailed evidence for the specific recognition forms participants proposed, including formal working time allocation, compensation for feedback contributions, and acknowledgment in role descriptions or performance evaluations.
&
The framework generalizes these conditional-willingness patterns into a design commitment centered on \emph{resourced voluntariness}: participation must remain optional, visibly recognized, and supported by protected time or equivalent organizational acknowledgment, rather than being treated as an informal extension of journalistic duties. The analytical move is from individual resistance to imposed burden and unpaid extra labor (an empirical finding about when participation is perceived as illegitimate) to an organizational design requirement that structures participation so that non-participation carries no professional penalty and contributions are formally recognized.\\
\midrule

P2: Continuous Value Alignment
&
Primarily grounded in the \emph{procedural agency} dimension of the perception gap (\ref{subsec:perception-gap}) and in Pattern~A, with reinforcement from Pattern~B and some overlap with accountability-related concerns later developed in P5.
&
Participants repeatedly framed participation as normative work concerning standards, rules, and accountability, explicitly distinguishing it from just improving a product or service [P4, P7, P2, P3]. P7 stated: "participation can't just be about the product or making it work better from a technical point of view. It has to be normative, regulatory. Journalists should be involved in setting the rules and must figure this out all together". P4 similarly emphasized the need for "clear guidelines about what's acceptable and how we keep our standards when using these tools". Section \ref{subsec:participation-meaning} further showed that internal governance participation concerned defining permissible use cases, disclosure requirements, verification standards, and acceptable boundaries of use. Pattern~B reinforces the processual character of this principle: participants valued training and capacity-building only when it was ongoing and institutionally embedded, integrated into working hours, connected to concrete newsroom policies, and sustained through on-the-job experimentation rather than treated as a one-off event [P7, P9]; a preference that directly parallels the logic of continuous rather than episodic alignment.
&
The framework extends these findings from a demand for normative voice into a design commitment to ongoing alignment: values, standards, and newsroom rules should not be treated as fixed once established, but as matters for periodic re-evaluation and collective revision. This step goes beyond the interview material in a design-oriented way. The analytical move is from journalists' insistence that participation should concern the \emph{substance} of norms (an empirical finding about what participation should be about) to a processual framework requirement in which value negotiation becomes an ongoing organizational practice rather than a one-off consultation event.\\
\midrule

P3: Contextual Knowledge Scaffolding
&
Primarily grounded in the \emph{epistemic controllability} dimension of the perception gap (\ref{subsec:perception-gap}) and in Pattern~B, with some overlap with legitimacy concerns where access to training was constrained by workload.
&
Participants emphasized the need for profession-specific knowledge about what AI can and cannot be relied upon for, including normative and practical boundaries [P1, P2]. P1 stated: "What I need are clear guidelines about what I must not do with AI specifically as a journalist. And management needs to commit to training employees". P2 wanted "specific courses on AI in journalism, not generic AI courses \ldots the regulatory and legal side, applied to our work and profession, specifically". P7 preferred practical, on-the-job experimentation ("trying things out with AI, understanding where it fails") over theoretical instruction, while P9 stressed that time scarcity made formal courses difficult to attend. Section \ref{subsec:participation-meaning} further documented that participants distinguished journalism-specific training from generic AI literacy and identified existing knowledge silos as barriers to collective participation.
&
The framework translates these findings into a design commitment to modular, situated, profession-specific scaffolding distributed across multiple learning modalities, including peer learning, tailored guidance, and on-the-job experimentation. The analytical move is from participants' expressed needs regarding training format, content, and accessibility (an empirical finding about the conditions under which epistemic access becomes possible) to a framework requirement that knowledge support be embedded within ongoing work practices and made available through multiple entry points, so that competence does not function as a gatekeeping prerequisite for participation.\\
\midrule

P4: Sovereignty Preservation
&
Primarily grounded in \emph{procedural agency}, with important overlap with \emph{epistemic controllability} where understanding AI behavior serves as a precondition for exercising editorial authority over its use.
&
Participants did not evaluate AI merely in terms of output quality, but through whether editorial sovereignty would be preserved and whether they could exercise informed professional judgment over AI-mediated work. Pattern~A most directly grounds this principle: participants wanted influence over "what is allowed", "who decides", and "what happens when something goes wrong", framing these not as interface-level preferences but as questions of editorial authority and decisional primacy. Section \ref{subsec:participation-meaning} further showed that internal governance participation included demands for clarity about accountability escalation when AI outputs raise editorial concerns. The epistemic controllability dimension reinforces these concerns from a complementary angle: participants asked for practical and normative boundaries, including recognizing failure modes and what the epistemic controllability theme describes as boundaries of acceptable reliance [P1, P2], suggesting that the ability to understand system behavior is a necessary condition for exercising meaningful editorial override.
&
The framework sharpens these concerns into a requirement that human editorial judgment must retain decisional primacy over AI outputs, with explicit override mechanisms and protected spaces for professional disagreement with the system. The analytical move is from journalists' situated insistence on retaining authority over AI-mediated work (an empirical finding about the conditions under which AI remains professionally acceptable) to a design requirement that systems and workflows encode human override, document divergences from AI suggestions, and preserve editorial primacy throughout implementation.\\
\midrule

P5: Transparent Accountability
&
Primarily grounded in \emph{procedural agency}, with overlap with \emph{epistemic controllability} where provenance and traceability concerns support the demand for audit mechanisms, and with sovereignty concerns where accountability and editorial control intersect.
&
Participants consistently linked meaningful participation to knowing who is responsible for decisions, how responsibilities are distributed, and what happens when AI-related failures occur. In Section \ref{subsec:perception-gap}, procedural agency was framed as influence over standards and accountability; in Pattern~A, participants raised questions about "who decides" and "what happens when something goes wrong". Section \ref{subsec:participation-meaning} further specified that internal governance participation included accountability escalation when AI outputs raise editorial concerns, attribution of responsibility, and disclosure requirements when AI is used in published work. The epistemic controllability dimension reinforces this principle from a complementary direction: participants' concerns about provenance, meaning where AI gets its information, what it can and cannot be relied upon for, and how to recognize failure modes [P1, P2], directly support the need for traceable audit mechanisms that make system behavior legible. Additionally, the external feedback participation arena (\ref{subsec:participation-meaning}) extended accountability demands beyond the newsroom: participants who envisioned reporting errors and limitations to vendors were articulating a form of cross-organizational traceability in which accountability structures span institutional boundaries.
&
The framework formalizes these concerns into transparent accountability mechanisms: traceable decision processes, clear attribution of responsibility, and disclosure structures that make governance visible to differently positioned stakeholders. The analytical move is from journalists' demand for traceability and responsibility in cases of error or contestation (an empirical finding about the conditions of legitimate governance) to a design requirement that decision processes themselves be rendered visible, auditable, and legible across the newsroom and, where relevant, beyond it.\\

\bottomrule
\end{longtable}

\section{Supplementary Material: Operationalizing GVP Depth and Scope}
\label{appendix:gvp_operationalization}

In this appendix, Table~\ref{tab:gvp_operationalization} provides an operationalization of the two dimensions of the GVP framework, \emph{depth} and \emph{scope}, through working definitions, observable indicators, typical decision rights, and example governance artifacts or practices. The table is intended to clarify how positions in the GVP matrix may be interpreted in practice across newsroom AI governance settings. In particular, it specifies what can count as participation at different levels of depth, from more preliminary forms of involvement to more durable infrastructuring arrangements, and across different levels of scope, from local newsroom practices to broader sectoral and public governance arenas.

The table should be read as a lightweight interpretive aid rather than as a rigid classificatory instrument. Its purpose is not to reduce participation to a fixed measurement scheme, but to make explicit the kinds of activities, forms of influence, and governance objects associated with different positions in the matrix. In this sense, it complements the checkerboard representation introduced in Section \ref{gvp_framework} by clarifying how participation can be located, compared, and discussed analytically. The operationalization draws on prior work on layered participation in foundation-model ecosystems and on multi-level tensions in journalism, which inform respectively the \emph{depth} and \emph{scope} dimensions adapted in GVP.

\begin{longtable}{p{0.10\linewidth}p{0.14\linewidth}p{0.19\linewidth}p{0.19\linewidth}p{0.16\linewidth}p{0.17\linewidth}}
\caption{Operationalization of GVP depth and scope levels through working definitions, observable indicators, decision rights, and example governance artifacts. The table adapts the \emph{scope} dimension from prior work on journalism’s macro/meso/micro tensions and organizational levels \cite{suresh_participation_2024}, and the \emph{depth} dimension from prior work on layered participation in foundation-model ecosystems, including journalism as a relevant case study \cite{tseng_ownership_2025}. GVP extends these foundations by specifying how participation can be located and compared in newsroom AI governance through concrete activities, artifacts, and forms of influence.}\label{tab:gvp_operationalization}\\\label{tab:gvp_operationalization}\\
\toprule
\textbf{Dimension} & \textbf{Level} & \textbf{Working definition} & \textbf{Observable indicators} & \textbf{Typical decision rights} & \textbf{Example artifacts / practices}\\
\midrule
\endhead

\textbf{Depth} & \textbf{Policy definition} & Stakeholders articulate concerns, values, red lines, and acceptable-use boundaries for AI in journalism. & Participation in policy discussions; comments on draft rules; identification of prohibited uses; requests for disclosure, attribution, or verification standards. & Can propose, contest, or review norms and acceptable-use rules, but does not directly shape workflow execution or technical implementation. & AI-use policy drafts; acceptable-use checklists; ethics memos; disclosure rules; newsroom charters on AI use.\\

\textbf{Depth} & \textbf{Process shaping} & Stakeholders influence how AI is incorporated into newsroom workflows, including review procedures, escalation paths, and accountability routines. & Input on verification steps; specification of override points; definition of escalation procedures; participation in training design; allocation of responsibilities across roles. & Can shape workflow conditions, review requirements, and accountability procedures surrounding AI use. & Workflow diagrams; verification checklists; escalation protocols; approval rules; training modules; override procedures.\\

\textbf{Depth} & \textbf{Co-design} & Stakeholders participate in iterating tool use, interaction patterns, prompts, interface requirements, or evaluation criteria at the level of situated use. & Structured feedback on tool behavior; collaborative testing; pilot participation; prompt or template refinement; development of evaluation rubrics for newsroom tasks. & Can materially influence the design or adaptation of tool use at the application layer, without necessarily affecting upstream model development. & Prompt libraries; pilot feedback forms; interface mock-ups; annotated error cases; evaluation rubrics; newsroom-specific usage templates.\\

\textbf{Depth} & \textbf{Infrastructuring} & Stakeholders shape durable sociotechnical arrangements that persist beyond a single tool or pilot and organize ongoing participation, accountability, or inter-organizational coordination. & Creation of shared repositories; establishment of audit structures; definition of procurement conditions; coordination of cross-newsroom standards; development of long-term governance routines. & Can shape persistent governance infrastructures, organizational memory, and inter-organizational conditions for AI adoption and oversight. & Audit-log schemas; vendor clauses or SLAs; shared repositories; cross-newsroom standards; governance boards; accreditation or certification mechanisms.\\

\textbf{Scope} & \textbf{Micro} & Participation concerns individual or team-level work practices within a newsroom. & Local workflow changes; desk-level use norms; role-specific training; team discussions about acceptable use. & Decisions affect a single role, desk, or team. & Desk-specific checklists; prompt templates; local guidance notes; team review routines.\\

\textbf{Scope} & \textbf{Meso} & Participation concerns organization-wide coordination or collaboration across multiple desks or newsrooms. & Shared policy formation; inter-desk standards; cross-functional meetings; newsroom-wide training; coordination across organizations. & Decisions affect the newsroom as an organization, or a collaboration among multiple newsrooms. & Organization-wide policies; shared training repositories; joint codes of conduct; common review procedures.\\

\textbf{Scope} & \textbf{Macro} & Participation concerns sectoral, professional, or industry governance beyond a single organization. & Input into professional guidelines; consultation through associations or unions; contribution to sector-wide standards; negotiation of shared governance arrangements. & Decisions affect professional norms, industry standards, or collective bargaining positions. & Professional charters; journalism-association guidance; industry standards; model contractual clauses.\\

\textbf{Scope} & \textbf{Public} & Participation includes readers, civic actors, or other public-facing accountability processes concerned with the legitimacy of AI use in journalism. & Reader feedback channels; visible provenance or disclosure; public consultation; contestation mechanisms; community oversight. & Decisions affect public-facing transparency, accountability, and legitimacy conditions. & Disclosure templates; provenance markers; public complaint channels; community review procedures; reader-facing audit summaries.\\

\bottomrule
\end{longtable}
\end{document}